\documentclass[aps,pre,reprint,superscriptaddress,showpacs]{revtex4-1}

\usepackage{amsmath}
\usepackage{amssymb}
\usepackage{upgreek}
\usepackage{graphicx}
\usepackage{dcolumn}
\usepackage{hyperref}

\renewcommand{\vec}[1]{\ensuremath{\mathbf{#1}}}
\newcommand{\unitvec}[1]{\vec{\hat{#1}}}

\newcommand{\Dt}{\ensuremath{\Delta t}}
\newcommand{\Dx}{\ensuremath{\Delta x}}
\newcommand{\Mach}{\ensuremath{\mathrm{Ma}}}
\newcommand{\Reynolds}{\ensuremath{\mathrm{Re}}}
\newcommand{\Womersley}{\ensuremath{\alpha}}
\newcommand{\Dean}{\ensuremath{\kappa}}

\newcommand{\ud}{\,\mathrm{d}}
\newcommand{\cs}{\ensuremath{c_\textrm{s}}}
\newcommand{\feq}{\ensuremath{f^{\textrm{eq}}}}

\newcommand{\PhantomSite}{\ensuremath{\mathcal{P}}}
\newcommand{\IoletSite}{\ensuremath{\mathcal{I}}}

\newcommand{\iolet}{inlet/outlet}

\newcommand{\HemeLB}{\textsc{HemeLB}}

\newcolumntype{d}[1]{D{.}{.}{#1}}
\newcommand{\tblhead}[1]{\multicolumn{1}{c}{#1}}
\newcommand{\spantwo}[1]{\multicolumn{2}{c}{#1}}

\begin{document}

\title{Choice of boundary condition for lattice-Boltzmann simulation
  of moderate Reynolds number flow in complex domains}

\newcommand{\CoMPLEX}{\affiliation{CoMPLEX, University College London,
    Physics Building, Gower Street, London, WC1E 6BT, United Kingdom}}

\newcommand{\CCS}{\affiliation{Centre for Computational Science,
    Department of Chemistry, University College London, 20 Gordon
    Street, London, WC1H 0AJ, United Kingdom}}

\newcommand{\RC}{\affiliation{Research Software Development Team,
    Research Computing and Facilitating Services, University College
    London, Podium Building - 1st Floor, Gower Street, London, WC1E
    6BT, United Kingdom}}

\author{Rupert W. Nash}\CCS
\author{Hywel B. Carver}\CCS\CoMPLEX
\author{Miguel O. Bernabeu}\CCS\CoMPLEX
\author{James Hetherington}\CCS\RC
\author{Derek Groen}\CCS
\author{Timm Kr\"uger}\CCS
\author{Peter V. Coveney}\email[Corresponding author: ]{p.v.coveney@ucl.ac.uk}\CCS

\date{\today}

\begin{abstract}
  Modeling blood flow in larger vessels using lattice-Boltzmann
  methods comes with a challenging set of constraints: a complex
  geometry with walls and inlet/outlets at arbitrary orientations with
  respect to the lattice, intermediate Reynolds number, and unsteady
  flow. Simple bounce-back is one of the most commonly used, simplest,
  and most computationally efficient boundary conditions, but many
  others have been proposed. We implement three other methods
  applicable to complex geometries (Guo, Zheng and Shi, Phys Fluids
  (2002); Bouzdi, Firdaouss and Lallemand, Phys. Fluids (2001); Junk
  and Yang Phys. Rev. E (2005)) in our open-source application
  \HemeLB{}. We use these to simulate Poiseuille and Womersley flows
  in a cylindrical pipe with an arbitrary orientation at
  physiologically relevant Reynolds (1--300) and Womersley (4--12)
  numbers and steady flow in a curved pipe at relevant Dean number
  (100--200) and compare the accuracy to analytical solutions. We find
  that both the Bouzidi-Firdaouss-Lallemand and Guo-Zheng-Shi methods
  give second-order convergence in space while simple bounce-back
  degrades to first order. The BFL method appears to perform better
  than GZS in unsteady flows and is significantly less computationally
  expensive. The Junk-Yang method shows poor stability at larger
  Reynolds number and so cannot be recommended here. The choice of
  collision operator (lattice Bhatnagar-Gross-Krook vs.\ multiple
  relaxation time) and velocity set (D3Q15 vs.\ D3Q19 vs.\ D3Q27) does
  not significantly affect the accuracy in the problems studied.
\end{abstract}


\maketitle

\section{Introduction}
In the last two decades, lattice-Boltzmann methods
(LBM)~\cite{Succi:2001,Chen:1998co,Aidun:2010} have been widely
studied and used for fluid flow problems.  We are particularly
interested in its application to the study of hemodynamics: the flow
behavior of blood under physiological conditions.  Accurate simulation
of the flow of blood in an individual could have near-term clinical
benefits for, \emph{inter alia}, the treatment of
aneurysms~\cite{Kung:2011do,Villa-Uriol:2010,He:2008ir} or
stenoses~\cite{Spilker:2007eq, Taylor:2009dd, Tahir:2011vr}.

Applications such as these have a number of challenges (retaining
computational performance in a relatively sparse, three-dimensional
(3D) fluid domain; capturing the complex rheology of a dense
suspension of deformable particles; accounting for the compliance of
vessel walls) but here we address the choice of boundary condition
method in complex domains. We will not examine the rheology and compliance problems in
this article, restricting ourselves to a Newtonian fluid within a
rigid-walled system, but instead aim to provide recommendations for
the choice of boundary condition method(s) for a complex
geometry. This choice is examined in the context of two other
variables: the discrete velocity set and lattice-Boltzmann collision
operator. Bearing in mind recent controversy over the reproducibility
of computational science~\cite{Merali:2010,Morin:2012}, we have
released the source code of our application \HemeLB{} for the
community's scrutiny and use.

L\"att \emph{et al.}~\cite{Latt:2008fl} considered five boundary
conditions and assessed their accuracy. However, they studied only
boundary conditions in which the wall passes through a lattice point,
immediately restricting their results to boundaries that are normal to
one of the Cartesian directions of the underlying grid. Boyd \emph{et
  al.}~\cite{Boyd:2004kr} compared simple bounce-back to the
Guo-Zheng-Shi method~\cite{guo:2007} in simulations of arterial
bifurcations, finding that the two methods give different results when
stenosis is present, but without any analysis suggesting which, if
either, is more accurate. Stahl \emph{et al.}~\cite{Stahl:2010ie}
performed LBM simulations with simple bounce-back boundary conditions
to examine the effect of staircased boundaries (where walls that are
not aligned with the underlying grid are approximated by a set of grid
edges) on the measurements of shear stress both at the wall and in the
bulk flow, finding errors in the shear stress of up to 35\% for
two-dimensional (2D) channel flow and 28\% for 2D bent-pipe flow. They
also introduced a method for estimating the local wall normal from the
shear stress tensor. Pan \emph{et al.}~\cite{Pan:2006hj} have studied
the effect of different boundary conditions in porous media for flows
at low Reynolds number, concluding that interpolation at the
boundaries significantly improves accuracy.

In this paper we comprehensively compare the accuracy of simulations
performed with some of the most widely used solid-fluid boundary
conditions against analytical solutions relevant for simulations of
blood flow in the larger vessels. The outline of the paper is as
follows. In Sec.~\ref{sec:latticeBoltzmann} we briefly introduce the
lattice-Boltzmann method and discuss the collision operators, velocity
sets and boundary conditions used.  Our open-source
software~\cite{hemelb-website} and simulation protocols are described,
and results given in Sec.~\ref{sec:sims}. We present our conclusions
in Sec.~\ref{sec:conclusion}.

\section{The lattice-Boltzmann method}\label{sec:latticeBoltzmann}

Here we provide a brief summary of the lattice-Boltzmann method (LBM);
for a full derivation, we refer the reader to one of the many thorough
descriptions available~\cite{Chen:1998co,Succi:2001,Aidun:2010}. LBM
operates at a mesoscopic level, storing a discrete-velocity
approximation to the one-particle distribution functions of the
Boltzmann equation of kinetic theory~\cite{Liboff:1998}, $\{f_i
(\vec{r}, t)\}$, on a regular lattice, with grid spacing $\Dx$. The
set of velocities $\{\vec{c}_i\}$ is chosen such that the distances
travelled in one timestep ($\Dt$), $\vec{e}_i = \vec{c}_i\Dt$, are
lattice vectors and to ensure Galilean
invariance~\cite{Qian:1992ke}. When one only wishes to reproduce
Navier-Stokes dynamics, the set is typically a subset of the Moore
neighborhood, including the rest vector. For 3D simulations, the most
commonly used sets have 15, 19 and 27 members (termed D3Q15, D3Q19 and
D3Q27, respectively). The LBM can be shown, through a Chapman-Enskog
expansion, to reproduce the Navier-Stokes equations in the
quasi-incompressible limit with errors proportional to the Mach number
squared.

In the absence of forces, the density $\rho (\vec{r}, t)$ and the
velocity $\vec{u}(\vec{r}, t)$ at a fluid site can be calculated from
the distributions by
\begin{eqnarray}
 \rho &=& \sum_i f_i, \\
 \rho \vec{u} &=& \sum_i f_i \vec{c}_i.
\end{eqnarray}

Advancing the system one timestep is conceptually divided into two
stages. The first is collision, which relaxes the distributions
towards a local equilibrium (we denote the post-collisional
distributions as $f_i^\star$):
\begin{equation}
 f_i^\star(\vec{r},t) = f_i(\vec{r},t) + \hat\Omega(f_i(\vec{r},t)),
 \label{eq:lb-general-relaxation}
\end{equation}
where $\hat\Omega$ is a the collision operator. The second is
streaming, where the post-collisional distributions are propagated
along the lattice vectors to new locations in the lattice, defining
the distributions of the next timestep:
\begin{equation}
 f_i(\vec{r} + \vec{c}_i \Dt, t + \Dt) = f_i^\star(\vec{r}, t).
\end{equation}

\subsection{Collision operators}\label{sec:collision-operators}

The collision operator approximates the microscopic interparticle
interactions. Here we summarize the two considered in this article:
lattice Bhatnagar-Gross-Krook and multiple relaxation time.

\textbf{Lattice Bhatnagar-Gross-Krook}
(LBGK)~\cite{Qian:1992ke,Chen:1992jj} approximates the collision
process as relaxation towards a local equilibrium (see below), in a
discrete velocity analogue of the Boltzmann-BGK approximation from
kinetic theory~\cite{Bhatnagar:1954},
\begin{equation}\label{eq:lbgk-relaxation}
  \hat\Omega(f_i) = - \frac{(f_i - \feq_i)}{\tau}\Dt,
\end{equation}
where $\tau$ is the relaxation time. In this case all modes relax
towards equilibrium at the same rate. This can be shown, through a
Chapman-Enskog expansion (see, e.g., \cite{Ladd:2001,Chen:1998co}), to
reproduce the Navier-Stokes equations with a kinematic viscosity $\nu$
given by
\begin{equation}
  \label{eq:lbgk-visc}
  \nu = \cs^2 (\tau - \Dt/2).
\end{equation}
For the equilibrium distribution, we use a second-order (in velocity
space) approximation to a Maxwellian distribution
\begin{equation}
 \feq_i (\rho, \vec{u}) = \rho w_i \left(1 + \frac{\vec{c}_i \cdot \vec{u}}{\cs^2} 
 + \frac{ (\vec{c}_i \cdot \vec{u})^2}{2 \cs ^ 4} 
 - \frac{\vec{u}\cdot\vec{u}}{2 \cs^2}\right),
\end{equation}
where the weights $w_i$ and speed of sound $\cs$ depend on the choice
of velocity set. LBGK is simple to implement and gives excellent
performance; it is therefore widely used.

\textbf{Multiple relaxation time} (MRT) collision operators, developed
at the same time as LBGK~\cite{dHumieres:1992vj}, generalize the
notion of relaxation towards local equilibrium (the same as above) by
allowing different relaxation rates for different moments of the
distributions, potentially improving stability properties and
accuracy~\cite{dHumieres:2002gq}. The eigenvalue of the collision
matrix which corresponds to the relaxation of shear stress
$\lambda_\textrm{shear}$ determines the viscosity as in
\eqref{eq:lbgk-visc} ($\tau\rightarrow 1/\lambda_\textrm{shear}$). We
use the MRT operator on the D3Q15 and D3Q19 lattices, as described by
d'Humi\`eres \emph{et al.}~\cite{dHumieres:2002gq}. Here, the
distribution function $f_i$ are transformed into the moment basis
$m_i$ by the matrix $M_{ij}$ and the different moments can be relaxed
towards equilibrium at different rates, before being projected back
into the distribution space for advection
\begin{equation}
  \hat\Omega(f_i) = -\sum_j M^{-1}_{ij} s_j (m_j - m^{\textrm{eq}}_j),
\end{equation}
where $s_j$ is the relaxation rate for mode $j$.

\subsection{No-slip boundary conditions}\label{sec:noslip-bc}

Boundary conditions for LBM have some additional complications
compared to boundary conditions for Navier-Stokes based methods, due
to the LBM's mesoscopic nature. Typically, one wishes to impose
conditions on the macroscopic, hydrodynamic variables ($p$,
$\vec{u}$) but these must be implemented through a closure relation
for the mesoscopic distributions. There is no single, obviously
superior choice. In this section, we will briefly review some commonly
used boundary condition methods for lattice-Boltzmann models which
impose the no-slip condition that the velocity of fluid adjacent to a
wall is equal to the velocity of the wall.

Many boundary condition methods do not vary their behavior with
respect to the location of the walls in relation to the Eulerian
grid. The wall is often assumed to pass infinitesimally close to a
point, the grid point remaining inside the fluid domain (sometimes
referred to as ``wet node'' boundary conditions~\cite{Latt:2008fl}),
or alternatively the boundary is considered to be halfway along the
lattice vector to a point outside the fluid. In the case of complex or
non-lattice-aligned domains, these methods (and simple bounce-back,
see below) will always cause a first-order \emph{modeling} error,
irrespective of the order of numerical accuracy of the resulting
lattice-Boltzmann method. (Typically second order accuracy is sought,
since this is the inherent accuracy of standard lattice-Boltzmann
methods in bulk fluids.) This point has been studied numerically by
Stahl \emph{et al.}~\cite{Stahl:2010ie}. Other boundary conditions
allow the wall to be at an arbitrary position along the link between a
solid and a fluid site. These reduce the modeling error of fixed wall
position methods, but often at the price of increased complexity
and/or the requirement of data from neighboring fluid lattice
points. Further, a number of methods suffer reduced accuracy at sites
in corners, which can reduce the accuracy of simulations throughout
the domain.

There are a number of popular methods~\cite{inamuro:2928, Zou:1997tp,
  Latt:2007, Ansumali:2002kr} that can only operate on straight,
axis-aligned planes and force the boundary to pass directly through
the lattice point. Malaspinas \emph{et al.}~\cite{Malaspinas:2011gb}
generalized the regularized method~\cite{Latt:2007} to cope accurately
with corner nodes. The authors acknowledge that it fails for the case
of a D3Q15 lattice and a right-angled corner and this method also
forces the boundary to pass through a lattice point. These methods
are, therefore, unsuitable for problems involving complex boundaries.

\textbf{Simple bounce-back (SBB)} is perhaps the most widely used
boundary condition for solid walls, positioning them halfway along the
lattice vector from fluid to solid. It is straightforward to implement
and gives second-order accurate simulation results for flat boundaries
aligned with the Cartesian axes of the lattice~\cite{Ziegler:1993},
although in more complex cases this degrades to first-order
accuracy~\cite{Ginzbourg:1994}. It is also computationally cheap and
local in its operation. SBB ensures conservation of mass up to machine
precision. In this work we use the halfway bounce-back
scheme~\cite{Ladd:1994a}. Despite SBB exhibiting the modeling error
mentioned above, we include it in this study due to its wide use.

The \textbf{Bouzidi-Firdaouss-Lallemand (BFL)}
method~\cite{Bouzidi:2001di} starts with simple bounce-back, but
interpolates the value of the to-be-propagated distribution with the
distribution at the fluid site which standard bounce-back would stream
it to. They present two variants: one using linear interpolation and
another using quadratic interpolation. In the present work, we
restrict our attention to the linear case, due to its locality and
smaller communication requirement (indeed, it can be implemented
without any inter-process communication above the normal
lattice-Boltzmann streaming step, albeit at a price of revisiting the
sites adjacent to the wall). Bouzidi \emph{et al.} claim that both
variants show second-order convergence, but with the linear method
having a poorer prefactor~\cite{Bouzidi:2001di}.  The method as
presented by Bouzidi \emph{et al.} may, depending upon the distance of
the wall from the fluid site, fail when computing $f_i(\vec{x},
t+\Dt)$ where $\vec{x}-\vec{e}_i$ lies outside the fluid domain, if
the site at $\vec{x}+\vec{e}_i$ is also outside. Since this can occur
in a curved domain, in these cases we fall back to performing
SBB. While this does degrade accuracy slightly, it does appear to
offer good stability of the simulation.

\textbf{Guo, Zheng, and Shi (GZS)}~\cite{guo:2007} present a boundary
condition which decomposes the unknown distributions at the wall into
equilibrium and non-equilibrium parts. The equilibrium part uses the
density of the fluid site and a linearly extrapolated velocity such
that the velocity at the solid wall is as imposed. For the
non-equilibrium part, the value from the fluid site (or the next site
into the bulk) is used. The method as described in~\cite{guo:2007} has
the same problem as BFL. We again fall back to SBB when this occurs.

\textbf{Junk and Yang (JY)}~\cite{Junk:2005cj} propose a correction to
the simple bounce-back method. They claim an advantage compared to
interpolation-based methods such as GZS and BFL as their method is
completely local and is able to handle non-straight boundaries where
sites have lattice vectors in opposite directions, both crossing the
solid-fluid boundary. The method arises from their
analysis~\cite{Junk:2005ea} of boundary conditions for LBM in terms of
general methods for studying finite difference schemes rather than the
standard Chapman-Enskog expansion. By adding correction terms to the
collision operator and then discretizing them in an optimal (under
their analysis framework) manner, they ensure the Navier-Stokes
equations are obeyed with the correct boundary conditions.

The multireflection method by Ginzburg and
d'Humi\`eres~\cite{Ginzburg:2003jr} uses linear combination of five
neighboring distribution functions along a link direction plus a
correction to determine the unknowm direction. This combination is
obtained from a Chapman-Enskog expansion at the boundary sites, which
is third-order accurate for steady flows. Chun and
Ladd~\cite{Chun:2007bd} present a method based only on interpolation
of the equilibrium part of the distribution functions, relying on the
fact that the non-equilibrium part is always one order higher in the
Chapman-Enskog expansion. Chun and Ladd demonstrate numerically that
their method is advantageous for problems with many small gaps, such
as porous media.

It is well known that LBGK combined with SBB does not locate the plane
of zero velocity exactly halfway along the link for flows with varying
velocity gradients~\cite{He:1997kb} and that this can be a significant
issue in, for example, porous media~\cite{Pan:2006hj}. The effective
width of the channel in lattice units $H$, for a Poiseuille flow in a
2D lattice-aligned channel, driven by a constant body force is given
by (Eq.~(19) from~\cite{Luo:2011ir})
\begin{equation}
H^2 = N_x^2 + 48 \nu^4 - 1
\end{equation}
where $N_x$ is the number of lattice points across the channel and
$\nu$ is the viscosity in lattice units.  While this error can be
reduced by optimizing the choice of MRT relaxation
times~\cite{dHumieres:2009dw}, it can only reach zero for particularly
simple cases. Typically at $\Reynolds \ge 1$ the larger system sizes
and smaller relaxation times combine to reduce the relative error. For
example with $N_x=40$ and $\tau=0.55$ the effective width is 39.987, a
relative error of $0.03\%$. Due to the smallness of these errors in
the problems studied here, we do not consider them further.

We also note that some authors propose the use of
grid-refinement~\cite{Filippova:1998}, finite
difference~\cite{He:1996jw}, and finite
volume~\cite{Chen:1998eq,Ubertini:2003dy} discretizations of the
discrete Boltzmann equation as methods for improving accuracy around
non-planar boundaries, but we will restrict our attention here to
implementations on a single, regular grid. Additionally, the immersed
boundary method (IBM)~\cite{Peskin:2002} has been used in conjunction
with the LBM to simulate rigid~\cite{Feng:2004} and
deformable~\cite{Zhang:2007wv} boundaries. IBM requires a further
layer of fluid sites outside the walls as well as another, simpler
boundary condition at the edge of the halo region, but admits an
extension to moving boundaries in a natural way.

\subsection{Open boundary conditions}\label{sec:open-bc}

For inlets, we use Ladd's method~\cite{Ladd:1994a} to impose the
expected velocity profile (parabolic, Womersley). This is a
modification of SBB, with a correction term $-2 w_i \rho
\vec{u}\cdot\vec{c}_i/c_s^2$, where $\vec{u}$ is the imposed velocity
at the halfway point of the link. However, using this at outlets as
well will cause the total mass in the system to increase unboundedly,
due to the unbalanced mass fluxes into and out from the system since
LBM has a finite compressibility~\cite{Kruger:2009ce}.

Alternatively, one could impose mixed Dirichlet-Neumann boundary
conditions at the outlet:
\begin{subequations}
  \begin{equation}\label{eq:pressure-bc}
    p = p_0,
  \end{equation}
  \begin{equation}\label{eq:para-velocity-bc}
    \vec{u}_\parallel = 0,
  \end{equation}
  \begin{equation}\label{eq:perp-velocity-bc}
    \unitvec{n}\cdot\nabla u_\perp = 0,
  \end{equation}
\end{subequations}
where $\unitvec{n}$ is the inward pointing normal of the open
boundary.

A number of authors~\cite{Zou:1997tp, Chikatamarla:2006tt,
  Mattila:2010dp, Mattila:2009to} have proposed open boundary
conditions that fulfill some or all of these requirements, however the
techniques cited are only suitable for inlets aligned with the
lattice's axes. We have therefore developed a simple method that meets
these requirements.

Assume that, at the start of a timestep, all distributions for an
inlet/outlet site are known. LBM then proceeds as normal: a collision
step, followed by a streaming step; the distributions that would have
come from exterior sites now have an undefined value. We close the
system by constructing a ``phantom site'' (indicated below with a
subscript $\PhantomSite$) beyond the boundary, whose hydrodynamic
variables are estimated based on the imposed values and those at the
inlet/outlet site (indicated with subscript $\IoletSite$). Note that
there is one phantom site per missing distribution, i.e., the phantom
sites of adjacent \iolet{} sites are unrelated. This is to eliminate
the need for communication between neighboring sites. The equilibrium
distribution for the missing direction is computed at the phantom site
and then streamed into place. For the density, we assume the target
pressure $p_0$, i.e., we perform a zero-order extrapolation from the
outlet plane. Condition \eqref{eq:para-velocity-bc} is enforced by
projecting away any velocity component not parallel to the \iolet{}
normal, $\unitvec{n}$. For condition \eqref{eq:perp-velocity-bc}, we
take first-order finite-difference approximations for the derivatives,
giving:
\begin{equation}
  \unitvec{n}\cdot\nabla u_\perp \approx
  \frac{u_\perp(\vec{r}_\IoletSite, t) -
    u_\perp(\vec{r}_\PhantomSite, t)}{c_{i \alpha}\cdot
    \unitvec{n}\Dt} = 0.
\end{equation}
Hence
\begin{equation}
  \vec{u}(\vec{r}_\PhantomSite, t) \approx
  (\vec{u}(\vec{r}_\IoletSite, t)\cdot\unitvec{n})\unitvec{n}.
\end{equation}
For lattice sites that are adjacent to both open and closed
boundaries, we use the above method on those links that cross the
\iolet{} and the solid wall boundary condition on those links that
cross the solid wall of the domain.


\section{Simulations}\label{sec:sims}

Our goal is to determine which combination of boundary condition,
collision operator and velocity set gives the best all-round accuracy
in a non-trivial geometry (i.e., with curved surfaces), with a focus on
computational hemodynamics. We also assess the computational
requirements of the different models.

We compare against analytical solutions, which restricts us to
relatively simple domains: we choose a cylinder and a torus. For the
cylinder we use both steady, Hagen-Poiseuille flow and a
time-dependent, sinusoidal Womersley flow. For the torus, we use only
steady flow. By choosing a non-axis-aligned orientation for the
cylinder, we better mimic a typical production simulation of the human
vasculature. The orientation $\unitvec{n}$ was chosen pseudorandomly
from the unit sphere, subject to the constraint that $\unitvec{n}
\cdot \unitvec{e}_i \le 0.9$, $\forall i$. The value is
\begin{equation}
  \unitvec{n} = [-0.299, 0.382, 0.874].
\end{equation}
Our approach is to select parameters in lattice units, but with
physiologically relevant Reynolds (\Reynolds) and Womersley
(\Womersley) numbers.

\subsection{Software: HemeLB}\label{sec:software}

The simulations in this paper were performed with
\HemeLB{}~\cite{Mazzeo:2008fd}, a lattice-Boltzmann-based fluids
solver, which includes capability for \textit{in situ} imaging of
flow-fields and real-time steering~\cite{Mazzeo:2010ji}. It is a
distributed memory application, parallelized with MPI. We have shown
that \HemeLB's computational performance scales linearly up to at
least 32,768 cores~\cite{Groen:2012wy}. We have released the software
online~\cite{hemelb-website}, under the open-source GNU Lesser General
Public License (LGPL), to enable interested researchers to reproduce
our results as well as to use the software for novel problems.

\HemeLB{} has several linked components, described in Groen \emph{et
  al.}~\cite{Groen:2012wy}. We have recently re-developed the
lattice-Boltzmann core to allow for easy switching between use of
different velocity sets, collision operators, and boundary conditions,
through a statically polymorphic, object-oriented design. This avoids
any run-time overhead due to dynamic
polymorphism~\cite{Alexandrescu:2001tj}. The individual software
components are tested through a battery of over one hundred unit and
regression tests, which are run nightly by our continuous integration
server.

\HemeLB{} includes, amongst other features: the D3Q15, D3Q19 and D3Q27
velocity sets; the lattice Bhatnagar-Gross-Krook (LBGK) and multiple
relaxation time (MRT) collision operators; and, the simple bounce-back
(SBB), Guo-Zheng-Shi (GZS), Bouzidi-Firdaouss-Lallemand (BFL), and
Junk-Yang (JY) boundary condition methods. HemeLB does not currently
support the combination of the GZS boundary condition with the MRT
collision operator; this will be addressed in a future release.

The software includes a separate tool for defining the simulation
domain. This requires either a geometric primitive or a general
surface, meshed with triangles. The user can then place inlets and
outlets, specify their pressure/velocity boundary conditions, and
select the fineness of the lattice. This setup tool then generates the
input for \HemeLB{} itself, producing a description of each fluid site
and, if needed, the location of the wall. For the cylinders used in
this work, we directly use the cylinder, rather than approximating it
with triangles.

\subsection{Convergence analysis}\label{sec:convergence}

In this section we report on a series of simulations of
Hagen-Poiseuille flow over a range of resolutions and Reynolds
numbers, defined here as
\begin{equation}\label{eq:reynolds}
  \Reynolds = \frac{U_\mathrm{max} D}{\nu},
\end{equation}
where $U_\mathrm{max}$ is the maximum velocity, $D$ the pipe diameter,
and $\nu$ the kinematic viscosity. The velocity $\vec{U}$ is
\begin{equation}
  \vec{U} = \frac{\partial_n p}{4 \rho\nu}\left(\left(\frac{D}{2}\right)^2 - r^2\right)\unitvec{n},
\end{equation}
where $r$ is the distance from the cylinder axis, defined by
$\unitvec{n}$, and $\partial_n p$ is the pressure gradient along the
axis.

\begin{table}[tb]
  \caption{
    Parameters for convergence analysis. All values are given in 
    lattice units. Parameters are, from left to right: the Reynolds
    number $\Reynolds$; the cylinder diameter $D$;
    the predicted Mach number $\Mach$; the LBGK relaxation time
    $\tau$; the relative density difference imposed
    $\Delta\rho/\rho_0$; the momentum diffusion time
    $T_\mathrm{mom}\equiv D^2/\nu$, and the sound propagation time $T_s\equiv L/\cs$.
    \label{tab:converge-params}
  }
  \begin{ruledtabular}
    \begin{tabular}{d{0}d{0}d{4}d{2}d{5}d{0}d{0}}
      \tblhead{$Re$} & \tblhead{$D$} & \tblhead{$\Mach$} & \tblhead{$\tau$} & \tblhead{$\Delta\rho/\rho_0$} & \tblhead{$T_\mathrm{mom}$} & \tblhead{$T_s$} \\
      \hline
      1 & 12 & 0.0241 & 1.00 & 0.01850 & 864 & 42 \\
      1 & 24 & 0.0120 & 1.00 & 0.00463 & 3460 & 83 \\
      1 & 48 & 0.0060 & 1.00 & 0.00116 & 13800 & 166 \\
      1 & 96 & 0.0030 & 1.00 & 0.00029 & 55300 & 333 \\
      \hline
      30 & 12 & 0.3610 & 0.75 & 0.13900 & 1730 & 42 \\
      30 & 24 & 0.1800 & 0.75 & 0.03470 & 6910 & 83 \\
      30 & 48 & 0.0902 & 0.75 & 0.00868 & 27600 & 166 \\
      30 & 96 & 0.0451 & 0.75 & 0.00217 & 111000 & 333 \\
      \hline
      100 & 12 & 0.4810 & 0.60 & 0.07410 & 4320 & 42 \\
      100 & 24 & 0.2410 & 0.60 & 0.01850 & 17300 & 83 \\
      100 & 48 & 0.1200 & 0.60 & 0.00463 & 69100 & 166 \\
      100 & 96 & 0.0601 & 0.60 & 0.00116 & 276000 & 333 \\
      \hline
      300 & 12 & 0.7220 & 0.55 & 0.05560 & 8640 & 42 \\
      300 & 24 & 0.3610 & 0.55 & 0.01390 & 34600 & 83 \\
      300 & 48 & 0.1800 & 0.55 & 0.00347 & 138000 & 166 \\
      300 & 96 & 0.0902 & 0.55 & 0.00087 & 553000 & 333 \\
    \end{tabular}
  \end{ruledtabular}
\end{table}
 
In Table~\ref{tab:converge-params} we list all the parameters chosen
for the simulations. The range of \Reynolds{} spans typical values for
cerebral arteries in the human body~\cite{bronzino2006BEF}. For each
case we vary the diameter $D$ from $12-192$ lattice units; the length
of tube used is given by $L = 4D$. Due to the finite speed of sound in
LBM ($\cs = 1/\sqrt{3}$ for the models used here), we list the Mach
numbers ($\Mach\equiv U_\mathrm{max} / \cs$); the lowest resolution
simulations in each have extremely high values and will consequently
have poor accuracy, but this allows us to assess the convergence
behavior at modest computational expense. Next we show the value of
the LBGK relaxation time $\tau$ which must be greater than
$\Dt/2$~\cite{Succi:2001} and not be much greater than
$\Dt$~\cite{Holdych:2004gi}. For the MRT simulations, we use the
parameters from~\cite{dHumieres:2002gq} except for the stress
relaxation rate where we use $s_9=s_{11}=1/\tau$, i.e., $s_1=1.6 \Dt$,
$s_2=1.2 \Dt$, $s_4=1.6 \Dt$ and $s_{14}=1.2 \Dt$.

We hold $\tau$ constant in lattice units
while refining the spatial resolution, which implies diffusive scaling
of the timestep (when converted to physical units), i.e., $\Dt \propto
\Dx^2$. The density, and hence pressure, difference $\Delta\rho$
driving the flow is also reported; this must remain much less than the
reference density of the simulation $\rho_0$ in order to keep
compressibility errors small. Finally, we list the time for momentum
to diffuse across the cylinder's diameter, $T_\mathrm{mom}\equiv
D^2/\nu$, and the time for a sound wave to propagate the length of the
cylinder, $T_s\equiv L/\cs$, to give some idea of the time required
for the simulation to converge to a steady state. To determine whether
a simulation has indeed converged, we compute the maximum difference
between flow fields at two times
\begin{equation}\label{eq:conv-measure}
  \Delta u(t_1, t_2) = \frac{\max_\vec{r}\|\vec{u}(\vec{r}, t_1) - \vec{u}(\vec{r}, t_2)\|}{U_\mathrm{max}},
\end{equation}
and require that $\Delta u(t, t+1) < 10^{-7}$.

We use a simple initialization procedure, initializing to a uniform
density fluid at rest. This approach is general and can be applied to
any geometry without requiring any preprocessing
step~\cite{Caiazzo:2005bn,Artoli:2006ha,Mei:2006tn}, but does require
longer simulation times until a steady solution is reached. For each
simulation we compare the Poiseuille solution, $\vec{U}(\vec{r})$,
with the velocity field found by simulation, $\vec{u}(\vec{r}, t)$. We
define the velocity error as
\begin{equation}\label{eq:u-star}
  \vec{u}^\star (\vec{r}, t)\equiv \vec{u}(\vec{r}, t) - \vec{U}(\vec{r})
\end{equation}
and use the $\ell^2$-norm scaled by the predicted velocity range as
our measure of error:
\begin{equation}\label{eq:vel-err-2}
  \epsilon^2_u(t) = \frac{\sqrt{\sum_\vec{r} {\vec{u}^{\star 2}}}}
  {\sqrt{N}\max_\vec{r}\|\vec{U}\|}.
\end{equation}
These are evaluated over all fluid sites in the central $90\%$ of the
cylinder, thus excluding the inlet and outlet sites. For the pressure
gradient, we use the measured difference at the edge of this volume
divided by the distance between the two planes. This is to disentangle
the errors due to the open boundary condition method from the no-slip
condition. We will return in the future to a full validation of the
open boundary condition method.

The simulations were performed on HECToR, the UK national
supercomputer, using up to 544 cores. The number of cores for each run
was chosen to minimize the run time while remaining efficient which we
have shown to occur at around $10^3$ sites per
core~\cite{Groen:2012wy}.

\begin{figure*}[htb]
  \centerline{
  \includegraphics{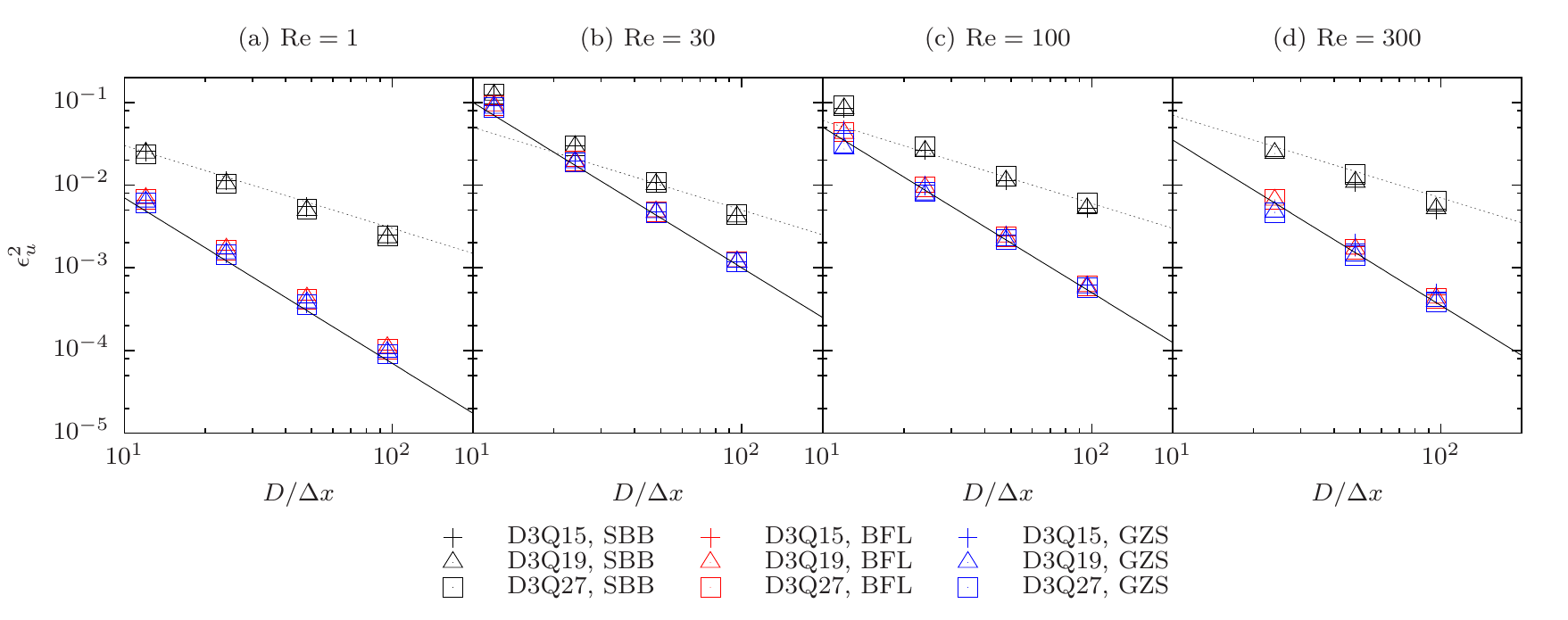}}
\caption{(Color online.) Convergence of velocity error residuals
  ($\epsilon^2_u$) as defined in Eq.~\eqref{eq:vel-err-2} vs.\
  diameter in lattice units. (a) Reynolds number $\Reynolds = 1$; (b)
  $\Reynolds = 30$; (c) $\Reynolds = 100$; (d) $\Reynolds = 300$.
  Colors (black, red, blue) indicate boundary condition (SBB, BFL,
  GZS) and shapes (crosses, triangles, squares) indicate the velocity
  set (D3Q15, D3Q19, D3Q27). The solid lines are guides to the eye
  showing first-order (dashed) and second-order (solid)
  convergence. The simulations with $\Reynolds=300$ and $D=12\Dx$
  became unstable due to the high Mach number and are not shown.
  \label{fig:converge-sbb}
  }
\end{figure*}

The Junk-Yang method showed poor stability (instability being defined
as one of the distribution functions becoming negative) and only
producing a converged solution for the lowest resolution cases at low
Reynolds number. We therefore disregard the JY method from further
consideration. We do however have some confidence in our
implementation of the method as our unit test suite demonstrates (data
not shown) that the implementation reduces to SBB when the wall is a
plane normal to one of the axes and halfway between the site and its
non-fluid neighbors, as expected~\cite{Junk:2005cj}.

In Fig.~\ref{fig:converge-sbb} we show the velocity errors as a
function of increasing resolution for the remaining three boundary
condition methods and the three velocity sets. We show only the
results for the LBGK collision operator (the results with MRT are
visually indistinguishable).

The results clearly show that SBB gives the expected first-order
convergence as the resolution increases, while BFL and GZS show
second-order convergence with similar prefactors. GZS offers slightly
superior accuracy than BFL, reaching up to 20\% lower errors
($\Reynolds=300$) but the relative benefit is highly variable,
vanishing in some cases.

As the Reynolds number is increased, we note that the errors also
increase, particularly when going from $\Reynolds = 1$ to $30$. This
is likely due to the increase in Mach number. The small improvement
from $\Reynolds=30$ to $300$ is probably due to the decreasing density
difference and hence reduced compressibilty errors.

The velocity set has a small effect on the measured accuracy: D3Q15
and D3Q19 are almost coincident. D3Q27 does offer a small benefit of
$1$--$4\%$ for GZS and BFL, but worsens accuracy for SBB; this may be
due to the greater number of sites which have to implement the
boundary condition and therefore suffer from the first-order modeling
error.

\subsection{Womersley flow}\label{sec:womersley}

Here we report on simulations of oscillatory flow, in order to explore
the different boundary condition methods' effect on accuracy for
time-dependent cases. The Womersley number (\Womersley) is a
dimensionless number governing dynamical similarity in cases of
oscillatory flow. It relates to the ratio of transient forces to
viscous forces (or alternatively the ratio of diameter to boundary
layer growth during one period of oscillation). In the case of a
cylinder (radius $R$, axis $\unitvec{n}$) and laminar flow with zero
average pressure gradient ($\partial_n p(t) = (A/L) \sin{\omega t}$,
$\omega\equiv 2\pi/T_\mathrm{osc}$), it is defined as:

\begin{equation}
  \label{eq:womersley}
  \Womersley = R\sqrt{\frac{\omega}{\nu}},
\end{equation}
where $\nu$ is the fluid viscosity. The time-dependent Navier-Stokes
equations admit an analytic solution~\cite{formaggia2009CM}:
\begin{equation}\label{eq:womersley-soln}
  \vec{u}(r, t) = 
  -\Re \left( 
    \frac{A}{\omega \rho L} 
    \left[
      1 - \frac{J_0(i^{3/2} \Womersley \frac{r}{R})}
      {J_0(i^{3/2} \Womersley)}
    \right] 
    e^{i \omega t}   
  \right)\unitvec{n},
\end{equation}
where $J_0$ is the order-0 Bessel function of the first kind and
$\Re(z)$ gives the real part of $z$.

\begin{table}[tb]
  \caption{
    Parameters for Womersley flow simulations. All values are given in
    lattice units. Parameters are: Reynolds number (defined for a
    Poiseuille flow with the maximum pressure gradient); Womersley
    number; predicted Mach number (based on the maximum Poiseuille
    flow); the LBGK relaxation time; the maximum relative density
    difference; the period of the pressure oscillation, and the
    momentum diffusion time.
    \label{tab:womersley-params}
  }
  \begin{ruledtabular}
    \begin{tabular}{d{0}d{0}d{3}d{4}d{5}d{0}d{0}d{0}}
      \tblhead{$Re$} & \tblhead{$\Womersley$} & \tblhead{$\Mach$} & \tblhead{$\tau$} & \tblhead{$\Delta\rho/\rho_0$} & \tblhead{$T_\mathrm{osc}$} & \tblhead{$T_\mathrm{mom}$} & \tblhead{$T_s$} \\
      \hline
      30 & 4 & 0.043 & 0.620 & 0.0040 & 5656 & 57600 & 333 \\
      100 & 8 & 0.078 & 0.565 & 0.0039 & 2608 & 107000 & 333 \\
      300 & 12 & 0.113 & 0.531 & 0.0027 & 2432 & 223000 & 333 \\
    \end{tabular}
  \end{ruledtabular}
\end{table}

In the larger arteries of the human body, Womersley numbers range
approximately between 4 and 20~\cite{bronzino2006BEF}, so we select
from this range. For these simulations, we define the Reynolds number
as that for a Poiseuille flow with a pressure gradient given by the
amplitude of the imposed gradient. Based on measured Reynolds and
Womersley numbers in human arteries~\cite{bronzino2006BEF}, we have
fit a simple power law $\Womersley = A \Reynolds ^ \gamma$, giving
$\gamma= 0.36$ and $A=0.1$. We select parameters only from this curve
within the $\Reynolds - \Womersley$ plane, as shown in
Table~\ref{tab:womersley-params}. We use the same cylinder orientation
as in Sec.~\ref{sec:convergence} and select a diameter $D = 48 \Dx$,
as this gives a reasonable balance between computational cost and
accuracy, such as would be chosen for production simulations; we keep
$L = 4D$. We initialize the simulation to a constant pressure with
zero velocity (with the phase of the pressure oscillation chosen such
that the driving difference is zero at simulation start). We allow the
simulation to run until $\Delta u(t, t-T_\mathrm{osc}) < 10^{-7}$ is
reached for all sample points during one oscillation; however, to
reduce the amount of data collected, we record data only for those
points within one lattice unit of an axis-normal plane halfway along
the cylinder. This was reached in 6 to 25 oscillation periods.

We run these simulations for \emph{all} combinations of LBM components
and compute residuals at four sample points during one pressure
oscillation period using Eq.~\eqref{eq:vel-err-2}. The four residuals
are then reduced by taking the root-mean-square average and the
maximum respectively, effectively extending the averaging/maximization
over time as well as space. We estimate the maximum as the maximum
velocity over a full period at the centre of the pipe, i.e.,
\begin{eqnarray}
  U_{\textrm{W}} = \frac{A}{\omega\rho L}&&
  \left\|1-\frac{1}{J_0(i^{3/2}\alpha)}\right\|\nonumber\\
 = \frac{4}{\alpha^2}&& \left\|1 - \frac{1}{J_0(i^{3/2}\alpha)}\right\|U_{\textrm{max}}
\end{eqnarray}
where $U_\mathrm{max}$ is the maximum velocity of a Poiseuille flow
driven by a pressure gradient of the amplitude of the Womersley flow.

\begin{figure}[tb]
  \includegraphics{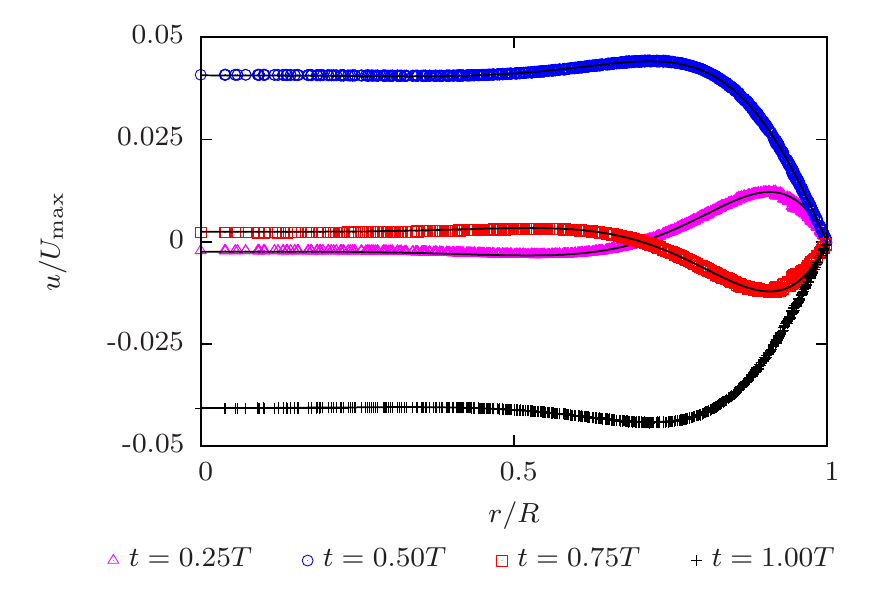}
  \caption{(Color online) Simulated (D3Q15, LBGK, BFL) axial velocity
    (normalized by $U_\mathrm{max}$) versus the radial coordinate. The
    Womersley number $\Womersley=12$ and the Reynolds number
    $\Reynolds=300$. The data is shown at four moments during the
    pulsatile cycle: $t=T/4$ (magenta, maximum positive pressure
    gradient), $T/2$ (blue, zero pressure gradient), $3T/4$, (red,
    maximum negative pressure gradient) $T$ (black, zero pressure
    gradient). The analytical solutions are shown with solids lines.}
  \label{fig:womersley-profile}
\end{figure}

\begin{table}[tb]
  \caption{Error residuals for pulsatile flow simulations with
    different LBM components. On the left are the boundary condition
    method (BC); collision operator (CO); and velocity set (DmQn). On
    the right are the error residuals $\epsilon^2_u$ for
    different values of Womersley number $\Womersley$ (other
    parameters are as shown in Table~\ref{tab:womersley-params}).
    \label{tab:womersley-results}
  }
  \begin{ruledtabular}
    \begin{tabular}{ccc|d{3}d{3}d{3}}
      & & & \multicolumn{3}{c}{$\epsilon^2_u$} \\
      BC & CO & DmQn & \tblhead{$\Womersley=4$} & \tblhead{$\Womersley=8$} & \tblhead{$\Womersley=12$} \\
      \hline
      SBB & LBGK & D3Q15 & 0.0085 & 0.0250 & 0.0343 \\
      SBB & MRT & D3Q15 & 0.0083 & 0.0248 & 0.0373 \\
      SBB & LBGK & D3Q19 & 0.0079 & 0.0234 & 0.0321 \\
      SBB & MRT & D3Q19 & 0.0090 & 0.0264 & 0.0395 \\
      SBB & LBGK & D3Q27 & 0.0089 & 0.0273 & 0.0395 \\
      BFL & LBGK & D3Q15 & 0.0013 & 0.0079 & 0.0128 \\
      BFL & MRT & D3Q15 & 0.0013 & 0.0078 & 0.0132 \\
      BFL & LBGK & D3Q19 & 0.0011 & 0.0069 & 0.0107 \\
      BFL & MRT & D3Q19 & 0.0012 & 0.0072 & 0.0118 \\
      BFL & LBGK & D3Q27 & 0.0012 & 0.0074 & 0.0124 \\
      GZS & LBGK & D3Q15 & 0.0013 & 0.0082 & 0.0139 \\
      GZS & LBGK & D3Q19 & 0.0011 & 0.0071 & 0.0115 \\
      GZS & LBGK & D3Q27 & 0.0012 & 0.0075 & 0.0128 \\
    \end{tabular}
  \end{ruledtabular}
\end{table}

In Fig.~\ref{fig:womersley-profile} we show an example of the
simulated velocities, for the D3Q15 velocity set, the LBGK collision
operator and the BFL boundary condition. We plot the axial velocity,
normalized by $U_\mathrm{max}$, against the radial coordinate at four
evenly spaced moments during the period of oscillation. The agreement
with the analytical profiles is excellent.

\begin{figure}[tb]
  \includegraphics{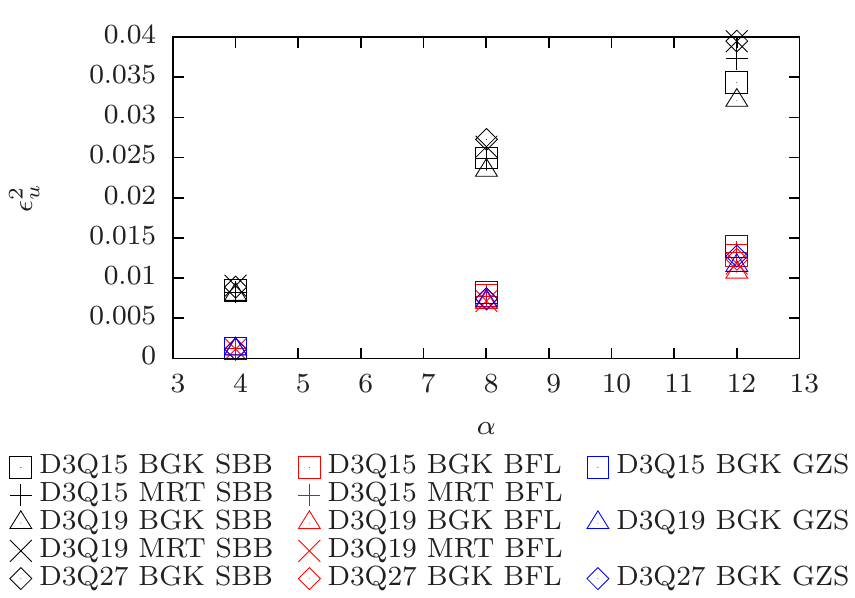}
  \caption{(Color online.) Error residuals $\epsilon^2_u$ as
    defined in Eq.~\eqref{eq:vel-err-2} for simulations of pulsatile
    flow vs.\ Womersley number $\Womersley$. Colors (black, red, blue)
    indicate boundary condition (SBB, BFL, GZS) and shapes the
    collision operator and velocity set. LBGK is indicated by polygons
    (square: D3Q15; triangle: D3Q19; diamond: D3Q27) and MRT by
    crosses ($+$: D3Q15; $\times$: D3Q19).
    \label{fig:womersley-error}
  }
\end{figure}

The $\ell^2$ error-norms for the simulations are shown in
Table~\ref{tab:womersley-results}. We see that all simulations well
approximate the analytical solutions, with errors in the range
$0.1$--$4\%$. In Fig.~\ref{fig:womersley-error}, we show the range of
error residuals against Womersley number. This clearly shows the the
BFL and GZS methods offers superior accuracy to SBB in all cases,
irrespective of the choice of collision operator and velocity
set. Choosing MRT over LBGK does not offer any benefits, however we
have not varied the relaxation rates for the non-stress tensor moments
of the distributions (e.g., by projecting out all the kinetic modes
every timestep or optimizing ``magic
numbers''~\cite{dHumieres:2009dw}). We note again that we have adopted
the parameters from~\cite{dHumieres:2002gq} without optimization. The
choice of velocity set does play a small role, especially in the
higher Reynolds and Womersley number cases, leading to slightly
improved results with larger velocity sets.

\subsection{Dean flow}\label{sec:dean}
\begin{figure}
\includegraphics{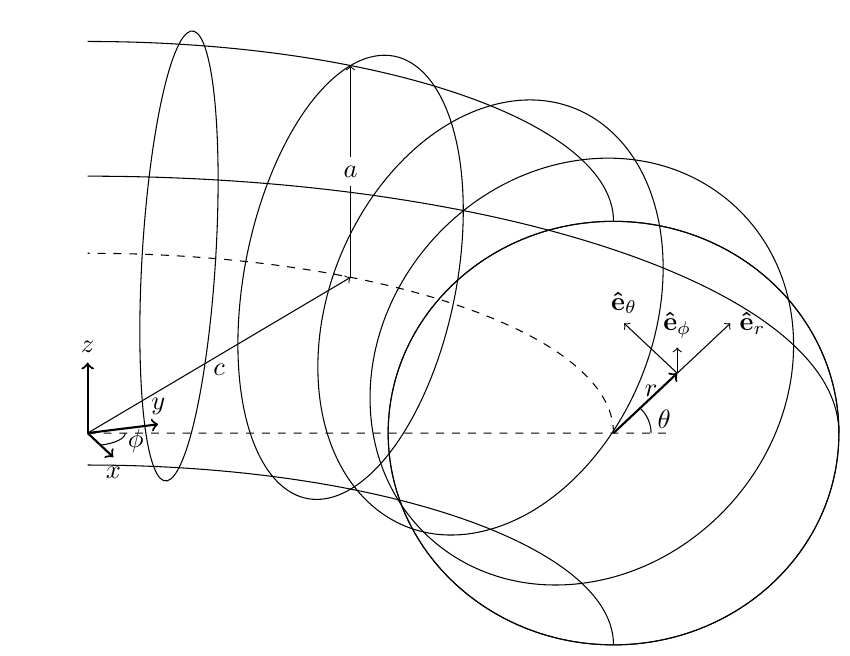}
\caption{Parameters and coordinate systems for the torus
  simulations. The radius of the tube is $a$ and the distance from the
  center of the tube to the center of the torus is $c$. We define a
  toroidal coordinate system $(r, \theta, \phi)$ where $(r, \theta)$
  are polar coordinates in the cross-section plane and $\phi$ is the
  angle that the cross-section plane makes with the $x$-$z$ plane.
}\label{fig:torus-gmy}
\end{figure}

\begin{figure*}[htb]
  \includegraphics[width=0.7\textwidth]{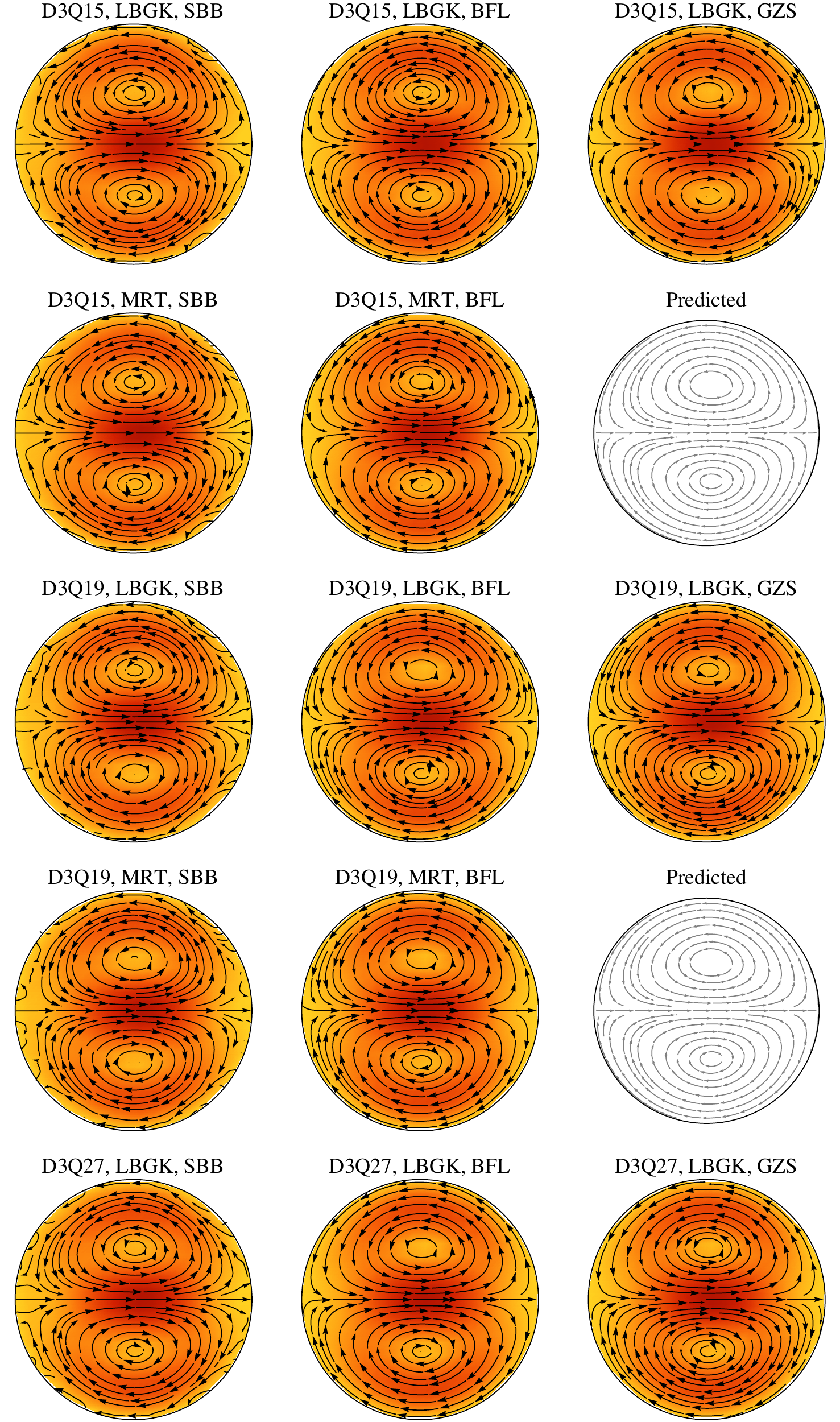}
  \caption{(Color online.) Streamlines for the secondary flow and
    color maps of the modulus of the error field in a torus, with Dean
    number $\Dean\approx 112$ and Reynolds number $\Reynolds\approx
    200$. The inside of the pipe is on the left of the figure. Across
    columns, we vary the boundary condition (SBB, BFL and GZS from
    left to right). Down the rows, we vary the combination of velocity
    set at collision operator (D3Q15+LBGK, D3Q19+LBGK, D3Q15+MRT,
    D3Q19+MRT, D3Q27+LBGK from top to bottom). The figures in color
    show the simulated results. The color field indicates the absolute
    error Eq.~\eqref{eq:dean-local-error} where light gray (yellow
    online) is zero and dark (red online) is $0.15 U_\phi(0,0)$. Since
    the combination of MRT and GZS is unavailable, we use these spaces
    to show the predicted secondary flow calculated using
    Eqs.~\eqref{eq:dean-uv-def}--\eqref{eq:dean-psi} in grayscale.}
  \label{fig:dean100}
\end{figure*}

The problems in the cylinders studied above are fundamentally 2D
flows. To more throroughly test the boundary conditions, we choose a
further benchmark problem that is fully 3D: flow in a torus. For
moderate Dean numbers (the dynamical similarity number for flow in
curved pipes; see below) the primary, axial flow is a perturbation of
a parabolic profile while the secondary flow in a cross sectional
plane is a pair of counter-rotation vortices---see the grayscale
images in Fig.~\ref{fig:dean100}. We define the radius of the tube as
$a$ and the distance from the center of the tube to the center of the
torus as $c$; our coordinate systems are illustrated in
Fig.~\ref{fig:torus-gmy}.

The characteristic number for flow in curved pipes is the Dean number
\Dean{} given by~\cite{Dean:1927eq,Dean:1928hj}
\begin{equation}
  \Dean = 4 \Reynolds \sqrt{2\delta}
\end{equation}
where \Reynolds{} is the Reynolds number based on the maxiumum axial
velocity of the flow that would result from the same pressure gradient
in a \emph{straight} pipe and $\delta\equiv a/c$ is the curvature
ratio. The Dean number is the square root of the product of the
inertial and centrifugal forces, scaled by the viscous
forces~\cite{Berger:1983kj}.

Dean~\cite{Dean:1927eq,Dean:1928hj} was the first to derive an
approximate solution for laminar, fully developed flow in a torus and
much of the later work is reviewed in~\cite{Berger:1983kj}. We choose
to use the approximations for a weakly curved pipe from Siggers and
Waters~\cite{Siggers:2005bi}, due to their clarity of
presentation. They define the velocity components $(U_r, U_\theta,
U_\phi)\equiv \frac{\nu}{a}(u,v,w)$ and the stream function $\psi$ by
\begin{equation}\label{eq:dean-uv-def}
u = \frac{1}{h r} \frac{\partial(h\psi)}{\partial\theta}
\mbox{ and }
v = -\frac{1}{h} \frac{\partial(h\psi)}{\partial r}
\end{equation}
where $h = 1 + \delta r \cos\theta$. They then expand as a power
series in \Dean{} 
\begin{equation}\label{eq:dean-soln}
  w = \Dean\sum_{k=0}^{\infty}\Dean^{2 k}w_k \mbox{ and } \psi = \Dean^2\sum_{k=0}^{\infty}\Dean^{2 k}\psi_k
\end{equation}
and each term $w_k$ and $\psi_k$ as a series in $\delta$:
\begin{equation}
  w_k = \sum_{j=0}^{\infty}\delta^{j}w_{kj} \mbox{ and } \psi_k = \sum_{j=0}^{\infty}\delta^{j}\psi_{kj}.
\end{equation}
We use only the terms
\begin{eqnarray}\label{eq:dean-w}
  w_{00} &=& \frac{1}{4}(1-r^2),\nonumber \\
  w_{01} &=& -\frac{3}{16}r(1-r^2) \cos\theta,\\
  w_{02} &=& \frac{1}{128}(1-r^2)(-3+11 r^2+10 r^2\cos 2\theta)\nonumber
\end{eqnarray}
and
\begin{eqnarray}\label{eq:dean-psi}
  \psi_{00} &=& \frac{1}{2^{10}\times 3^2} r (1-r^2)^2(4-r^2) \sin\theta,\nonumber\\
  \psi_{01} &=& \frac{1}{2^{12} \times 3^2\times 5} r^2 (1-r^2)^2(56-17 r^2) \sin 2\theta, \\
  \psi_{02} &=& \frac{1}{2^{17}\times 3^2\times 5}
  r (1-r^2)^2 [
    -(133-976 r^2 \nonumber \\
      &&+327 r^4) \sin\theta + 2 r^2\left(499-172 r^2\right) \sin3\theta
  ].\nonumber
\end{eqnarray}
This expansion is accurate for $\delta\ll 1$ and $\Dean \lesssim
100$~\cite{Siggers:2005bi}.

For our simulations, we select a curvature $\delta=0.1$ and two
\emph{target} Dean $\Dean_0$ numbers (100 and 200), as representative
of physiological values while not making the series expansion too
inaccurate. The tube radius $a=24\Dx$ (hence $c=240\Dx$) and the shear
relaxation time $\tau=0.6\Dt$ for $\Dean_0=100$ and $\tau=0.55\Dt$ for
$\Dean_0=200$. We sample the flow at a plane half way around the torus
to keep data volumes small and impose a parabolic flow profile at the
inlet and a constant pressure at the outlet. Since the imposed profile
is incorrect, we simulate 90\% of the full torus to allow distance for
the flow to fully develop.  One must also compute actual Dean number,
by equating the imposed flux
\begin{equation}
  Q_\mathrm{Pois}= \frac{\pi a\nu\Reynolds{}_0}{4} = \frac{\pi a \nu \Dean_0}{16 \sqrt{2\delta}}
\end{equation}
to the flux computed from Eq.~\eqref{eq:dean-soln}:
\begin{eqnarray}
  Q_\Dean{} &=& \int_0^1\int_0^{2\pi}  U_\phi a^2r\ud  r\ud\theta\nonumber \\
  &=& \frac{ \pi a \nu \Dean}{384} (48 + \delta^2),
\end{eqnarray}
giving:
\begin{equation}
\Dean= \frac{12 \Dean_0}{48+\delta^2}\sqrt{\frac{2}{\delta}}.
\end{equation}
This gives our actual Dean numbers as $\Dean\approx 111.78$ and
$\Dean\approx223.56$. We use the $\ell^2$-norm error from
Eq.~\eqref{eq:vel-err-2} but also evaluate it for the primary flow and
the seconday flow.

\begin{figure*}[htb]
  \includegraphics[width=0.7\textwidth]{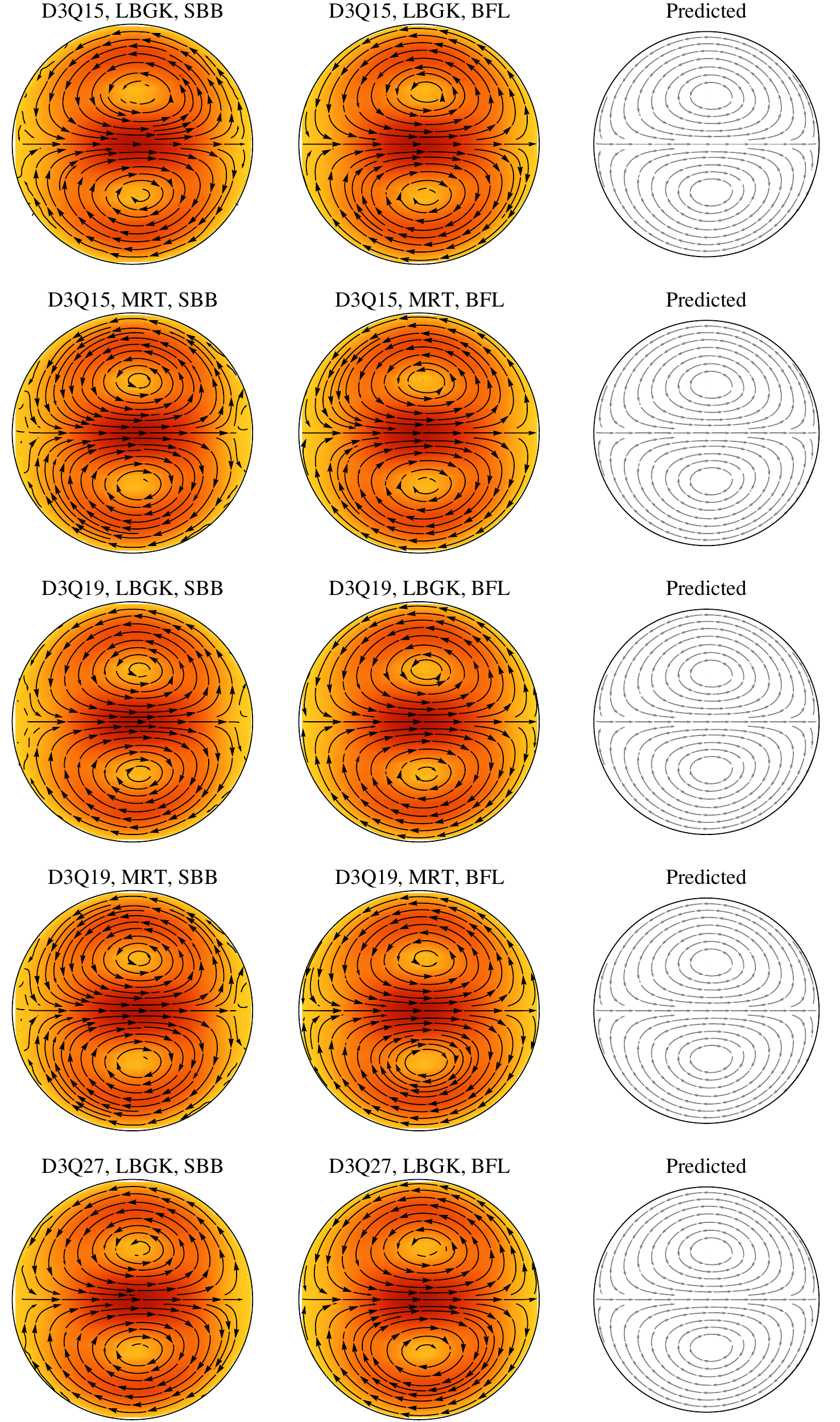}
  \caption{(Color online.) Streamlines for the secondary flow and
    color maps of the modulus of the error field in a torus, with Dean
    number $\Dean\approx 224$ and Reynolds number $\Reynolds\approx
    400$. The inside of the pipe is on the left of the figure. Across
    columns, we vary the boundary condition (SBB, BFL and GZS from
    left to right). Down the rows, we vary the combination of velocity
    set at collision operator (D3Q15+LBGK, D3Q19+LBGK, D3Q15+MRT,
    D3Q19+MRT, D3Q27+LBGK from top to bottom). The figures in color
    show the simulated results. The color field indicates the absolute
    error Eq.~\eqref{eq:dean-local-error} where light gray (yellow
    online) is zero and dark (red online) is $0.3 U_\phi(0,0)$. Since
    the GZS simulations were unstable, we use these spaces to show the
    predicted secondary flow calculated using
    Eqs.~\eqref{eq:dean-uv-def}--\eqref{eq:dean-psi} in grayscale.}
  \label{fig:dean200}
\end{figure*}

The simulations using the GZS boundary condition were unstable for the
larger \Dean/\Reynolds{} case, but the remaining results are collected
in Table~\ref{tab:dean-results}, and in Fig.~\ref{fig:dean100} and
Fig.~\ref{fig:dean200} we show streamline plots of the secondary flow,
colored by the modulus of the local absolute error
$\|\vec{u}^\star\|=\|\vec{u}-\vec{U}\|$ from Eq.~\eqref{eq:u-star},
scaled by $U_\phi$ evaluated at the tube center, i.e.,
\begin{equation}\label{eq:dean-local-error}
  \epsilon_\Dean=\frac{\|\vec{u}-\vec{U}\|}
  {U_\phi(0,0)}.
\end{equation}
The table of error norms shows that the error is dominated by the
error in the secondary flow, which is approximately independent of the
LBM model used. This is borne out by Fig.~\ref{fig:dean100} and
Fig.~\ref{fig:dean200} which show the patterns of the secondary flow
error (i.e., the coloring) remaining near-constant. The $\Dean\approx
224$ case has much larger errors that vary little which we ascribe to
the series solution becoming inaccurate.

We note that our simulations do locate the centers of the pair of
vortices accurately. The streamlines for the SBB simulations show
large errors in the velocity field's direction near the walls, due to
the stair casing of the boundary. This is likely to cause the stress
near the walls, which is a key hemodynamic factor, to have larger
errors. The primary flow errors for the case $\Dean\approx 112$ are
approximately $25\%$ larger for SBB than for either BFL or GZS, again,
due to the staircasing of the boundary. The velocity set and collision
operator do not strongly affect the measured error, but there is a
small benefit to using MRT over LBGK and for using the larger velocity
sets.

\newcommand{\spanthree}[1]{\multicolumn{3}{c}{#1}}
\begin{table*}[tb]
  \caption{Error residuals for flow in a toroidal pipe with
    different LBM components. On the left are the velocity set (DmQn)
    and collision operator (CO). On the right are the error residuals $\epsilon^2_u$ for
    different boundary collision methods for the secondary flow
    ($\epsilon^2_\mathrm{plane}$), the primary flow
    ($\epsilon^2_\mathrm{axial}$) and the overall
    ($\epsilon^2_\mathrm{all}$).
    \label{tab:dean-results}
  }
  \begin{ruledtabular}
    \begin{tabular}{cc|d{3}d{3}d{3}|d{3}d{3}d{3}|d{3}d{3}d{3}}
      & & \spanthree{SBB} & \spanthree{BFL} & \spanthree{GZS} \\
      DmQn & CO &
      \tblhead{$\epsilon^2_\mathrm{plane}$} & \tblhead{$\epsilon^2_\mathrm{axial}$} & \tblhead{$\epsilon^2_\mathrm{all}$} &
      \tblhead{$\epsilon^2_\mathrm{plane}$} & \tblhead{$\epsilon^2_\mathrm{axial}$} & \tblhead{$\epsilon^2_\mathrm{all}$} &
      \tblhead{$\epsilon^2_\mathrm{plane}$} &
      \tblhead{$\epsilon^2_\mathrm{axial}$} &
      \tblhead{$\epsilon^2_\mathrm{all}$} \\
      \hline
      & & \multicolumn{9}{c}{$\Dean\approx 112$, $\Reynolds\approx 200$} \\
      \hline
      D3Q15 & LBGK & 
      0.076 & 0.022 & 0.079 &
      0.076 & 0.018 & 0.078 &
      0.076 & 0.018 & 0.079 \\

      D3Q15 & MRT &
      0.076 & 0.021 & 0.079 &
      0.076 & 0.018 & 0.078 &
      \spanthree{-} \\

      D3Q19 & LBGK &
      0.075 & 0.020 & 0.078 &
      0.076 & 0.017 & 0.078 &
      0.076 & 0.017 & 0.078 \\
      
      D3Q19 & MRT &
      0.075 & 0.021 & 0.078 &
      0.076 & 0.017 & 0.078 &
      \spanthree{-}\\
      
      D3Q27 & LBGK &
      0.076 & 0.022 & 0.079 &
      0.076 & 0.017 & 0.078 &
      0.076 & 0.017 & 0.078 \\

      \hline
      & & \multicolumn{9}{c}{$\Dean\approx 224$, $\Reynolds\approx 400$} \\
      \hline
      D3Q15 & LBGK &
      0.153 & 0.058 & 0.164 &
      0.152 & 0.059 & 0.163 &
      \spanthree{-} \\
      
      D3Q15 & MRT &
      0.153 & 0.057 & 0.164 &
      0.152 & 0.059 & 0.163 &
      \spanthree{-} \\

      D3Q19 & LBGK &
      0.152 & 0.059 & 0.163 &
      0.152 & 0.058 & 0.163 &
      \spanthree{-} \\

      D3Q19 & MRT &
      0.151 & 0.060 & 0.163 &
      0.152 & 0.058 & 0.162 &
      \spanthree{-} \\

      D3Q27 & LBGK &
      0.152 & 0.059 & 0.163 &
      0.152 & 0.058 & 0.163 & 
      \spanthree{-} \\
      
    \end{tabular}
  \end{ruledtabular}
\end{table*}


\subsection{Relative performance}\label{sec:perf}

The relative performance of the options is germane to the choice of
which combination of lattice-Boltzmann components to use. In
Table~\ref{tab:perf} we give the measured per-core performance for the
lowest \Reynolds{}/\Womersley{} pulsatile flow simulations. All these
runs were performed on HECToR, a Cray XE6 supercomputer with two
16-core AMD Opteron 2.3~GHz Interlagos processors per node. The
simulations used 64 cores. The performance is given in millions of
site updates per second (MSUPS) and is based on timings that include
only the lattice-Boltzmann updates. We also estimate the relative time
to perform one site update for the different methods, assuming that an
SBB site update takes the same time as a bulk fluid site update. The
simulation domain includes 347,401 sites of which 42,340 (12\%) are at
the solid-fluid boundary and hence use the various boundary condition
methods.

\begin{table}[tb]
  \caption{Performance per core for different combinations of boundary
    condition, collision operator and velocity set. The columns
    indicate the combination of collision operator and velocity set,
    while the rows indicate the boundary condition. Omitted
    simulations are indicated with a dash (-). The values are given in
    millions of site updates per second (MSUPS) followed by the time
    to perform one boundary condition site update, relative to SBB, in
    parentheses.
    \label{tab:perf}
  }
  \begin{ruledtabular}
    \begin{tabular}{lrrrrrrrrrr}
      & \multicolumn{6}{c}{LBGK} & \multicolumn{4}{c}{MRT} \\
      & \spantwo{D3Q15} & \spantwo{D3Q19} & \spantwo{D3Q27} & \spantwo{D3Q15} & \spantwo{D3Q19} \\
      \hline
      SBB & 1.5 & (1)   & 1.2 & (1)   & 0.8 & (1)   & 0.8 & (1)   & 0.5 & (1) \\
      BFL & 1.4 & (1.6) & 1.2 & (1.4) & 0.8 & (1.3) & 0.8 & (1.3) & 0.5 & (1.2) \\
      GZS & 1.0 & (4.9) & 0.8 & (5.1) & 0.5 & (6.1) & \spantwo{-} & \spantwo{-} \\
    \end{tabular}
  \end{ruledtabular}
\end{table}

SBB gives the best computational performance in all cases while BFL
requires between $20\%$ and $60\%$ more compute time; the extra time
needed remains approximately constant across all collision operators
and velocity sets, but reduces as a fraction of the time to perform a
bulk site update. GZS requires approximately five times the
computational effort compared to a bulk site.

For the three velocity sets, the number of distribution updates per
second (i.e., $\mathrm{SUPS} \times Q$) remains approximately
constant---D3Q15: $22.5$; D3Q19: $22.8$; D3Q27: $21.6$. The cost of
changing the collision operator from LBGK to MRT is large, decreasing
performance by a factor of two. We do not place much emphasis on this
result as the MRT collision operator is implemented na\"ively and
there is scope to improve the performance.

\section{Conclusion}\label{sec:conclusion}

The majority of benchmark problems reported in the lattice-Boltzmann
literature use lattice-aligned geometries, rather than the complex
domains required by many applications. We have performed a comparison
between LBM simulation solutions, from our open source application
\HemeLB{}~\cite{hemelb-website}, and analytical solutions in a
non-lattice-aligned, curved domain up to Reynolds numbers of 300, with
steady and unsteady flow. We have varied the resolution of the grid
used and the different components of the algorithm (collision
operator, velocity set, no-slip boundary condition). We find that at
these moderate values of $\Reynolds$, the choice of velocity set and
collision operator do not greatly affect accuracy or stability, but
that the choice of no-slip boundary condition method is critical.

Recent studies~\cite{White:2011cp,Kang:2012vi} have shown that the
commonly used D3Q15 and D3Q19 velocity sets can give strongly
orientation (with respect to the underlying lattice) dependent results
when simulating flows for Reynolds numbers $\ge 250$, while the D3Q27
velocity set maintains good isotropy. The problems studied are in
relatively complex cases (a circular pipe with a narrowing; circular
and square pipes at higher Reynolds number) but they are comparable to
our simulations of flow in a toroidal pipe due to the strong
three-dimensionality of the Dean flow. We do not see any lattice
artifacts in these simulations, even with $\Reynolds\approx 400$ and
the strong secondary flows. Even so, one must be aware of the
possibility of anisotropy when simulating a complex flow and ensure
that it does not occur in one's work.

The Junk-Yang method shows poor stability and is therefore unsuitable
for our hemodynamic applications or other applications that require
even moderate Reynolds numbers.

Simple bounce-back (SBB) shows first-order convergence over a wide
range of resolutions and Reynolds numbers, as expected. We confirm
that the Bouzidi-Firdaouss-Lallemand (BFL) and (modified)
Guo-Zheng-Shi (GZS) methods both show second-order convergence over a
wide range of resolutions and Reynolds numbers. For the steady flow
problem, GZS has lower errors than BFL, by a variable amount (up to
20\%). For the time-dependent problem, BFL has lower errors than GZS, by
around 10\%. GZS also became unstable for the largest Reynolds number
flows in curved pipes.

The typical goal of most computational modeling is to solve the
desired system of equations to some problem-dependent accuracy in the
minimum time. Considering the steady flow problem as a proxy, we can
estimate which of the GZS and BFL boundary conditions will give the
shortest simulation time. Taking an average increase in accuracy for
GZS of $\sim10\%$, allows us to estimate that the GZS method can use a
$\sim 5\%$ lower resolution to obtain the same accuracy. Due to the
diffusive scaling of the timestep, the GZS method will require fewer
site updates than BFL by a factor of $\sim0.95^5\approx 0.79$ or
$21\%$ less. However the BFL method is more computationally efficient,
by $\sim 30\%$, for the typical surface:bulk ratios studied here, so
the equivalent-accuracy BFL simulation will complete sooner

We therefore recommend BFL as the best all-round boundary condition of
those tested, due to to the good accuracy, performance, stability and
simplicity of implementation. The GZS method also offers good accuracy
(in our modified form) but has worse stability at high Reynolds number
and much poorer performance. This last point may not matter in domains
with a relatively low number of wall boundary sites. SBB is much less
accurate, but may be acceptable for use in undemanding applications
and when development time is at a premium. We believe that these
results will prove helpful to the community when selecting methods for
simulating hemodynamics and comparable applications with
lattice-Boltzmann methods.

\begin{acknowledgements}
  This work made use of HECToR, the UK’s national high-performance
  computing service. We also thank Dr J\"org Sa\ss{}mannshausen for
  local computational support. This work was supported by the British
  Heart Foundation, EPSRC grants ``Large Scale Lattice Boltzmann for
  Biocolloidal Systems'' (EP/I034602/1) and 2020 Science
  (http://www.2020science.net/, EP/I017909/1), and the EC-FP7 projects
  CRESTA (http://www.cresta-project.eu/, grant no. 287703) and MAPPER
  (http://www.mapper-project.eu/, grant no. 261507).
\end{acknowledgements}

\bibliography{physics_validation}

\end{document}